\documentclass[twoside]{article}
\usepackage{amssymb,amsmath}
\usepackage{qic,epsfig}

\usepackage[ansinew]{inputenc}

\usepackage{mathrsfs}
\newcommand{\R}{\mathbb {R}}
\newcommand{\N}{\mathbb{N}}

\newtheorem{thm}{Theorem}
\newtheorem{lem}{Lemma}
\newtheorem{cor}{Corollary}
\newtheorem{con}{Conjecture}
\newtheorem{remark}{Remark}
\newtheorem{example}{Example}

\def\bajo#1#2{\smash{\mathop{#1}\limits_{#2}}}  %%%% una flecha, una igualdad, puntos, etc. (#1).

\begin{document}
\setlength{\textheight}{8.0truein}    %FOR 2ND PAGE ONWARDS

\runninghead{Variance of the sum of independent quantum computing errors}
            {J. Lacalle, L.M.~Pozo~Coronado}

\normalsize\textlineskip
\thispagestyle{empty}
\setcounter{page}{1}

%\copyrightheading{Vol.}{No.}{Year}{Page Nos.}
\copyrightheading{0}{0}{2003}{000--000}

\vspace*{0.88truein}

\alphfootnote

\fpage{1}

\centerline{\bf
VARIANCE OF THE SUM}
\vspace*{0.035truein}
\centerline{\bf OF INDEPENDENT QUANTUM COMPUTING ERRORS}
\vspace*{0.37truein}
\centerline{\footnotesize
%%%%%%%%%%%%%%%%%%%%%%%%%%%%%%%%%%%%
%put authors' name and address here
%%%%%%%%%%%%%%%%%%%%%%%%%%%%%%%%%%%%
J. Lacalle}
\vspace*{0.015truein}
\centerline{\footnotesize\it Dpto. de Matem\'atica Aplicada a las TIC, ETSI de Sistemas Inform\'aticos}
\baselineskip=10pt
\centerline{\footnotesize \it Universidad Polit\'ecnica de Madrid}
\centerline{\footnotesize \it C/ Alan Turing s/n, 28031, Madrid, Spain}
\vspace*{10pt}
\centerline{\footnotesize
L.M.~Pozo~Coronado}
\vspace*{0.015truein}
\centerline{\footnotesize\it Dpto. de Matem\'atica Aplicada a las TIC, ETSI de Sistemas Inform\'aticos}
\baselineskip=10pt
\centerline{\footnotesize \it Universidad Polit\'ecnica de Madrid}
\centerline{\footnotesize \it C/ Alan Turing s/n, 28031, Madrid, Spain}
\vspace*{0.225truein}
\publisher{(received date)}{(revised date)}

\vspace*{0.21truein}

%% \abstracts{first paragraph}{second paragraph}{third paragraph}
%% If there is only one paragraph, just keep the second and third empty
%% like the following one
\abstracts{
The sum of quantum computing errors is the key element both for the estimation and control of errors in quantum computing and for its statistical study. In this article we analyze the sum of two independent quantum computing errors, $X_1$ and $X_2$, and we obtain the formula of the variance of the sum of these errors:
$$
V(X_1+X_2)=V(X_1)+V(X_2)-\frac{V(X_1)V(X_2)}{2}.
$$
We conjecture that this result holds true for general quantum computing errors and we prove the formula for independent isotropic quantum computing errors.
}{}{}

\vspace*{10pt}

\vspace*{3pt}
\communicate{to be filled by the Editorial}

\vspace*{1pt}\textlineskip    %) USE THIS MEASUREMENT WHEN THERE IS
   %) A SECTION HEADING
%\vspace*{-0.5pt}
%\noindent
%%%%%%%%%%%%%%%%%%%%%%%%%%%%%%%%

\section{Introduction}

It is well known that the main challenge to achieve an efficient quantum computation is the control of quantum errors~\cite{Ga}. To address this problem, two fundamental tools have been developed: quantum error correction codes~\cite{CS,St1,Go1} in combination with fault tolerant quantum computing~\cite{Sh,Pr1,St2,Go2}.

In this article we study the sum of independent quantum computing errors. In order to achieve this goal, we represent $n-$qubits as points of the unit real sphere of dimension $d=2^{n+1}-1$, $S^d=\{x\in\R^{d+1}\ |\ \|x\| =1\}$, in coordinates with respect to the computational basis $[|0\rangle,|1\rangle,\dots,|2^n-1\rangle]$,
\begin{eqnarray}\label{QubitFormula}
\Psi=(x_0+ix_1,x_2+ix_3,\dots,x_{d-1}+ix_{d})
\end{eqnarray}
(see, for instance, \cite{NC})

We consider quantum computing errors as random variables with density function defined on $S^d$. It is easy to move from this representation to the usual representation in quantum computing through density matrices. Thereby, if $X$ is a quantum computing error with density function $f(x)$, then the density matrix of $X$, $\rho(X)$, is obtained as follows, using the pure quantum states given by formula (\ref{QubitFormula}):
$$
\rho(X)=\int_{S^d}f(x)|\Psi\rangle\langle\Psi|dx
\quad\text{where}\quad\int_{S^d}f(x)dx=1.
$$

It is well known that density matrices do not always discriminate different quantum computing errors (see \cite{NC}). Therefore, representations of quantum computing errors by random variables are more accurate than those by density matrices. This is the main reason why we have decided to use random variables to represent quantum computing errors. And, once the representation of quantum computing errors is established by random variables, the most natural parameter to measure the size of quantum computing errors is the variance.

The variance of a random variable $X$ is defined as the mean of the quadratic deviation from the mean value $\mu$ of $X$, $V(X)=E[\|X-\mu\|^2]$. In our case, since the random variable $X$ represents a quantum computing error, the mean value of $X$ is the $n-$qubit resulting from computing without error, $\Psi_0$. Without loss of generality, we will assume that the mean values of all quantum computing errors will always be $\Psi_0=|0\rangle$. For this, it is enough to move $\Psi_0$ into $|0\rangle$ through a unitary transformation. Therefore, using the pure quantum states given by formula (\ref{QubitFormula}), the variance of $X$ will be:
$$
V(X)=E[\|\Psi-\Psi_0\|^2]=E[2-2x_0]=2-2\int_{S^d}x_0f(x)dx.
$$

All quantum computing errors can be described in the framework that we have established above. Consequently, the formula that we have obtained for the variance of the sum of two independent quantum computing errors, $X_1$ and $X_2$, is a key result for the analysis of the propagation of errors in quantum computing. Said formula reads
\begin{eqnarray}\label{VarFormula}
V(X_1+X_2)=V(X_1)+V(X_2)-\frac{V(X_1)V(X_2)}{2}.
\end{eqnarray}

We found this formula by analyzing the behavior of quantum correction codes against a quantum computing error (random variable) $X$ with the following distribution function (in the particular case of two qubits): $$
f(\theta_0)=\frac{(2q-2)!!}{(2\pi)^q}\frac{(1-\sigma^2)}
{(1+\sigma^2-2\sigma\cos(\theta_0))^q},
$$
(see \cite{LPF}). Here, $\theta_0$ is the first polar angle at $S^d$, $q=2^n$ and $\sigma$ is a parameter belonging to the interval $(0,1)$, related to the variance by the formula $V(X)=2(1-\sigma)$.

To better understand the physical meaning of formula (\ref{VarFormula}) we have to
consider that, since $S^d$ is a compact variety of $\R^{d+1}$, the
variance cannot behave in the natural way it does in $\R^{d+1}$.
The variance of the sum of two independent errors can no longer be
the sum of the variances. However, an expansion of the range of
natural variances in $S^d$, the interval $[0,2)$, linearizes
formula (\ref{VarFormula}). In fact, the function
$$ f:[0,2)\to
[0,\infty)\quad\text{such that}\quad
f(x)=-\log\left(1-\frac{x}{2}\right)$$
transforms formula (\ref{VarFormula}) into a
linear formula for the linear variance $V_l(X)=f(V(X))$:
$$
V_l(X_1+X_2)=V_l(X_1)+V_l(X_2). $$
From the physical point of view,
it is very significant that the function $f$ and its inverse
$f^{-1}(x)=2\left(1-e^{-x}\right)$ are linked to the logarithmic
and exponential functions respectively. This fact makes, for example, that the
function $f^{-1}(x)$ remembers the exponential decay of quantum coherence.

Our main contributions in this article are formula (\ref{VarFormula}), which allows us to calculate the variance of the sum of two independent quantum computing errors in terms of their variances, and the following conjecture, which we are going to prove in the case of isotropic quantum computing errors.

\begin{con}
\label{Conjetura}
Let $X_1$ and $X_2$ be two independent quantum computing errors in a $n-$qubit computation, with variances $V_1(X_1)$ and $V_2(X_2)$ respectively. Then the variance of the sum of the errors, $V(X_1+X_2)$, verifies formula (\ref{VarFormula}).
\end{con}

To relate our error measure, variance, to the most common error measure in quantum computing, fidelity (see \cite{NC}), we define a quantum variance that takes into account that quantum states are equivalent under multiplication by a phase. The quantum variance of a random variable $X$ is defined as:
$$
V_q(X)=E[\bajo{\text{min}}{\phi}(\|\Psi-e^{i\phi}\Psi_0\|^2)]
=2-2E\left[\sqrt{x_0^2+x_1^2}\right].
$$

The fidelity of the random variable $X$, $F(X)$, with respect to the pure quantum state $\Psi_0=|0\rangle$ verifies $F(X)=\sqrt{\langle \Psi_0|\rho(X)|\Psi_0\rangle}$ (see \cite{NC}). Therefore, using the pure quantum states given by formula (\ref{QubitFormula}), $F(X)^2=E[\langle \Psi_0|\Psi\rangle\langle \Psi|\Psi_0\rangle]=E[|\langle \Psi_0|\Psi\rangle|^2]=E[x_0^2+x_1^2]$. Now, Jensen's inequality $\sqrt{E[x_0^2+x_1^2]}\geq E\left[\sqrt{x_0^2+x_1^2}\right]$ and the fact that $\sqrt{x_0^2+x_1^2}\geq x_0^2+x_1^2$ allow us to conclude that:
$$
1-\dfrac{V_q(X)}{2}\leq F(X)\leq\sqrt{1-\dfrac{V_q(X)}{2}}.
$$

These inequalities show that quantum variance and fidelity are essentially equivalent, since when quantum variance tends to $0$ fidelity tends to $1$ and, conversely, when fidelity tends to $1$ quantum variance tends to $0$.

In the quantum computing error measures considered so far, the null error corresponds to the values $V(X)=0$, $V_q(X)=0$ and $F(X)=1$, and their limit values are the following:
$$
0\leq V(X)\leq 4,\quad 0\leq V_q(X)\leq 2\quad\text{and}\quad 0\leq F(X)\leq 1
$$
Of the three measures, the one that best reflects the distribution of the $n-$qubit error is the variance. For this reason the goal of this article is to study this measure of error when two independent errors are added.

The variance, although its maximum value is $4$, tends to $2$ when we add independent errors (see table \ref{TablaVarianzas} below). Values greater than $2$ indicate that the distribution is concentrated on the hemisphere opposite the quantum state without error, $\Psi_0$. Therefore, these kind of distributions will not arise in normal situations. The qualitative analysis of the error accumulated in a quantum computation requires knowing the form of the distribution of the error, in addition to the value of the variance. There are many distributions with variance $2$ but we believe that the distribution to which the accumulated error distributions will tend is the uniform distribution over $S^d$, which of course also has variance $2$. If the accumulation of errors produces a variance close to $2$ and the limit distribution is the uniform distribution, the final $n-$qubit can be in any state with identical probability. This would mean that all information that the $n-$qubit may have contained would have been lost along the computation. We discuss this result in more detail in the conclusions.

\begin{table}[h]
\begin{center}
\label{TablaVarianzas}
\begin{tabular}{|c|c|c|}
\noalign{\hrule}
           & $\displaystyle 0<V_1<2$                          & $2<V_1<4$                            \\
\noalign{\hrule}
 $\displaystyle 0<V_2<2$ & $\text{max}\{V_1,V_2\}<V_{12}<2$   & $2<V_{12}<\text{min}\{V_1,4-V_2\}$   \\
\noalign{\hrule}
 $\displaystyle 2<V_2<4$ & $2<V_{12}<\text{min}\{4-V_1,V_2\}$ & $\text{max}\{4-V_1,4-V_2\}<V_{12}<2$ \\
\noalign{\hrule}
\end{tabular}
\vspace*{1pt}
\end{center}
\tcaption{Variance behavior according to formula (\ref{VarFormula})}
\end{table}

Random variables have been studied extensively in $\R^m$ for any dimension $m$ and, particularly, the formula for the variance of the sum of independent random variables has also been treated. In this sense, given errors (random variables) $X_1$,\dots $X_k$ in $\R^m$, it is well known that if they are independent then the variance verifies
$$
V(X_1+\cdots+X_k)=V(X_1)+\cdots+V(X_k)
$$

This statement is called the Bienaym\'e formula and was discovered in 1853 (see \cite{Bie}). Nevertheless, practically nothing is known if the errors are defined in $S^d$, including the formula (\ref{VarFormula}) that is presented for the first time in this article; surely because this analysis is very complicated and there have not been important applications so far that needed to be studied from this point of view.

This approach fits, for example, the study of errors in the position of a robotic arm whose components are connected by joints allowing rotational motion. However, the sensors that detect the position of the robotic arm allow a very effective control of the error in its position, making a deep analysis of these errors unnecessary.

As we have seen above, the behavior of errors in quantum computing also requires the use of random variables in spheres and, in this case, the laws of quantum mechanics do not allow any self-correcting system like that of sensors in robotic environments (within the code space if we are using quantum error correction codes). Therefore the development of the theory of random variables in $S^d$ has become essential to address the challenge of quantum computing and we think that the work presented in this article is an important achievement in this way.

There are many works related to the control of quantum computing errors, in addition to those already mentioned above. General studies and surveys on the subject~\cite{Pr2,Go3,CDT,PR,HFWH,AZ,DP,Fo,PFW}, about the quantum computation threshold theorem~\cite{AGP,SDT,WFSH,CT,WFHH}, quantum error correction codes~\cite{OG,LZLX,SS,Ha1}, concatenated quantum error correction codes~\cite{Fu} and articles related to topological quantum codes~\cite{DPB}. Lately, quantum computing error control has focused on both coherent errors~\cite{GM,BEKP} and cross-talk errors~\cite{PSVW,BTS}. Finally, we cannot forget the most complicated error to control in quantum computing, the quantum decoherence~\cite{Zu}. As we said above, these quantum computing errors can be analyzed in the framework of random variables on $S^d$ that we have established. This provides us with a general framework and allows us to apply the obtained results to any type of quantum computing error. Specifically, assuming that conjecture \ref{Conjetura} is true, formula (\ref{VarFormula}) allows us to study the behavior of the error (variance) during a quantum computation, by simply estimating the variance of the errors that accumulate in the process and their independence. In the conclusions we analyze in more detail the characteristics of the different types of error from the point of view of their control.

The outline of the article is as follows. In section 2 we set up notations and discuss some basic properties. In section 3 we include the main results: proofs of formula (\ref{VarFormula}) for the sum of independent isotropic quantum computing errors and for independent general errors in the case of dimension one, and an example to illustrate conjecture \ref{Conjetura} in the general case of independent quantum computing errors. Finally, in section 4 we explain the behavior of the variance of the accumulation of independent quantum computing errors, we discuss the properties of the distribution of accumulated errors and we analyze the characteristics of the different types of quantum computing error within the framework of the results presented in this article.

\vspace*{1pt}\textlineskip

\section{Notations and basic properties}

The results presented in sections 2 and 3 apply to both $n-$qubits quantum computing errors represented as random variables over $S^d$ with $d=2^{n+1}-1$, $S^d=\{x\in\R^{d+1}\ |\ \|x\| =1\}$ and $n-$qubits, in coordinates with respect to the computational basis $[|0\rangle,|1\rangle,\dots,|2^n-1\rangle]$,
$$
\Psi=(x_0+ix_1,x_2+ix_3,\dots,x_{d-1}+ix_{d})
$$
and error distributions over $S^d$ with arbitrary dimension $d\geq 1$, $S^d=\{x\in\R^{d+1}\ |\ \|x\| =1\}$ and coordinates
$$
x=(x_0,x_1,x_2,x_3,\dots,x_{d-1},x_{d})
$$

In the study of isotropic errors we will use the polar coordinates described below:
$$
\begin{array}{l}
0\leq\theta_0,\dots,\theta_{d-2}\leq\pi \\
-\pi\leq\theta_{d-1}<\pi
\end{array}\
\text{and}
\
\begin{array}{l}
x_j=\sin(\theta_0)\cdots\sin(\theta_{j-1})\cos(\theta_j),\ 0\leq j\leq d-1 \\
x_d=\sin(\theta_0)\cdots\sin(\theta_{d-1})
\end{array}
$$

From now on, we will consider only random variables (errors) $X$ on $S^d$ centered on the quantum state $\Psi_0=|0\rangle$, if we are working in quantum computing, and on the point $P=(1,0,\dots,0)$, if we are working with errors in spheres. In both cases, point $P$ represents the mean value of the error distribution. This assumption does not imply a loss of generality, because we can always transfer the mean to the point $P$ by a transformation of $U((d+1)/2)$ or $O(d+1)$. We will also always work in polar coordinates and will define the random variables by means of density functions $f(\theta_0,\theta_1,\dots,\theta_{d-1})$.

Henceforward we will write r.v. instead of random variable and r.vs. instead of random variables.

The variance of a r.v. $X$ is the mean of its quadratic deviation, i.e. $E[\|X-P\|^2]=E[(x_0-1)^2+x_1^2+\cdots+x_{d-1}^2+x_d^2)]=E[2-2x_0]$. And, in polar coordinates, the variance is $E[2-2\cos(\theta_0)]$.

\begin{definition}
The \emph{variance} of a r.v. $X$, centered on the point $P$, is $V(X)=E[2-2\cos(\theta_0)]$.
\end{definition}

Note that the variance will take values between $0$, if the density function is a delta at point $P$, and $4$, if the density function is a delta at point $-P$.

\begin{definition}
A r.v. in $S^d$ is \emph{isotropic} if its density functions depends exclusively on $\theta_0$ if $d\geq2$, and its values in $\theta_0$ and $-\theta_0$ are equal, if $d=1$.
\end{definition}

For simplicity, we will assume that the density function of any isotropic r.v. depends on $\cos(\theta_0)$. Indeed $g(\theta_0)=f(\cos(\theta_0))$, being $f$ the function $g\circ\arccos$ ($\cos$ is bijective in $[0,\pi]$).

The {\sl reasonable density functions} of isotropic r.vs. (errors) are not increasing with respect to the variable $\theta_0$. This means that the error probability (density function) does not grow when the error does. In this case, the variance is upper bounded by $2$ and takes value $2$ when the density function is constant in $S^d$. For example, in quantum computing this means that all information that the $n-$qubit may have has been completely lost.

In order to analyze the variance of the sum of independent isotropic r.vs. we need to obtain the expression of the resulting density function. Let $X_1$ and $X_2$ be independent isotropic r.vs. with density function $f_1$ and $f_2$ respectively and let $X=X_1+X_2$ and $f$ its density function.

The sum of r.vs. (errors) occurs as follows: initially there is no error and, therefore, our system is in $P$, when the first error occurs the system changes to $x^\prime$, with a probability that depends on $\|x^\prime-P\|^2$, and when the second error occurs the system goes to $x$, with a probability that depends on $\|x^\prime-x\|^2$ (isotropic error with respect to $x^\prime$). Then it holds
\begin{eqnarray}\label{FuncionDensidadGeneral}
f(x)=\int_{S^d}f_1\,f_2\,dx^\prime
\end{eqnarray}

\begin{thm}\label{thm1}
Given two independent isotropic r.vs. in $S^d$, $X_1$ and $X_2$, then the sum $X=X_1+X_2$ is also an isotropic r.v. in $S^d$.
\end{thm}
\begin{proof}
{Given $x_0^\prime$ and $x_0$, let us consider $S^\prime$ and $S$, the $d-1$ dimensional spheres of radius $1-x_0^{\prime\,2}$ and $1-x_0^2$ centered on $(x_0^\prime,0,\dots,0)$ and $(x_0,0,\dots,0)$ respectively. $S^\prime$ and $S$ are ``parallel" of $S^d$ passing through $x_0^\prime$ and $x_0$ respectively, if we consider $P$ as the $S^d$ polar point. Note that for $d=1$, $S$ and $S'$ are just pairs of points. By the symmetry of $S^\prime$ and $S$ and the isotropy of $f_1$ and $f_2$ we can conclude that the integral (\ref{FuncionDensidadGeneral}), restricted to $S^\prime$, is constant in $S$. Therefore the integral (\ref{FuncionDensidadGeneral}) is also constant in $S$ and, consequently, the sum $X$ of the two independent isotropic r.vs. is an isotropic r.v. in $S^d$}
\end{proof}

For the next corollary we mark that, as we shall state later in lemma~\ref{surface}, the surface of a $0$-sphere is $|S^0|=2$.

\begin{cor}\label{sum}
Given two independent isotropic r.vs. in $S^d$  ($d\geq 2$), $X_1$ and $X_2$, then the density function of the sum $X=X_1+X_2$ is
$$
f(\theta_0)=|S^{d-2}|\int_0^\pi\int_0^\pi f_1(\alpha)\,f_2(\beta)\,\sin^{d-1}(\theta_0^\prime)\sin^{d-2}(\theta_1^\prime)\,d\theta_0^\prime\, d\theta_1^\prime
$$
where $|S^{d-2}|$ is the surface of $S^{d-2}$, $f_1$ and $f_2$ the density functions of $X_1$ and $X_2$ respectively, $\alpha=\cos(\theta_0^\prime)$ and $\beta=\cos(\theta_0)\cos(\theta_0^\prime)+\sin(\theta_0)\sin(\theta_0^\prime)\cos(\theta_1^\prime)$.
\end{cor}
\begin{proof}
{The only questions that must be studied in the corresponding multiple integral are the values of $\alpha$ and $\beta$. As we have seen before in this section: $2-2\alpha=\|x^\prime-P\|^2=2-2\cos(\theta_0^\prime)$ implies $\alpha=\cos(\theta_0^\prime)$ and $2-2\beta=\|x^\prime-x\|^2$. By theorem~\ref{thm1}, we can choose $x=(\cos(\theta_0)+i\sin(\theta_0),0,\dots,0)$, if we are working in quantum computing, or $x=(\cos(\theta_0),\sin(\theta_0),0,\dots,0)$, if we are working with errors in spheres, and in both cases get $\|x^\prime-x\|^2=2-2(\cos(\theta_0)\cos(\theta_0^\prime)+\sin(\theta_0)\sin(\theta_0^\prime)\cos(\theta_1^\prime))$, completing the proof}
\end{proof}

\begin{remark}\label{rmk25}
For dimension $d=1$, the formula for the density function of the sum of two independent isotropic r.vs. in $S^1$, $X_1$ and $X_2$, is easily seen to be
\[
f(\theta_0)=\int_{-\pi}^\pi f_1(\alpha)\,f_2(\beta)\,d\theta_0^\prime,
\]
where $f_1$ and $f_2$ are the density functions of $X_1$ and $X_2$ respectively, $\alpha=\cos(\theta_0^\prime)$ and $\beta=\cos(\theta_0)\cos(\theta_0^\prime)+\sin(\theta_0)\sin(\theta_0^\prime)$.
\end{remark}

\vspace*{1pt}\textlineskip

\section{Variance of the sum of independent random variables}

\vspace*{1pt}\textlineskip

\subsection{Isotropic random variables}

We first state some technical results that will be used later.
\begin{lem} \label{intcossin}
We have%
\[
\int_{0}^{\pi}\cos^{a}\theta\sin^{b}\theta d\theta=\left\{
\begin{array}
[c]{l}%
\pi\frac{(a-1)!!(b-1)!!}{(a+b)!!}\quad\text{for }a\text{,}b\text{ even,}\\
2\frac{(a-1)!!(b-1)!!}{(a+b)!!}\quad\text{for }a\text{ even, }b\text{ odd,}\\
0\qquad\text{for }a\text{ odd}%
\end{array}
\right.
\]
(where $(-1)!!=1$).
\end{lem}

\begin{lem}\label{surface}
The surface of a $2d$-sphere (resp. $2d-1$) is%
\[
\left\vert S^{2d}\right\vert =\frac{2\left(  2\pi\right)  ^{d}}{(2d-1)!!}%
,\quad\left\vert S^{2d-1}\right\vert =\frac{\left(  2\pi\right)  ^{d}%
}{(2d-2)!!}
\]

\end{lem}
%
%Let us consider two isotropic error operators $\Phi_{1},\Phi_{2}$ given by
%probability density functions $f_{1},f_{2}$ defined on a $(d-1)$-sphere
%$S^{d-1}$ %%(note that the isotropy implies that $f_{i}$ depends only on the
%%%distance to the selected origin).
%The variance of $\Phi_{i}$ is given by%
%
%\[
%I_{i}=Var(\Phi_{i})=2-2\left\vert S^{d-2}\right\vert \int_{0}^{\pi}\cos
%\theta_{0}f_{i}(\cos\theta_{0})\sin^{2d-2}\theta_{0}d\theta_{0},
%\]
%while the variance of the sum of the two errors, applied sequentially, is
%given by%
%
%\begin{align*}
%I_{Sum}  & =Var(\Phi_{1},\Phi_{2})=2-2\left\vert S^{d-2}\right\vert \left\vert
%S_{d-3}\right\vert \int_{0}^{\pi}\int_{0}^{\pi}\int_{0}^{\pi}\cos\theta
%_{0}f_{1}(\cos\theta_{0}^{\prime})\\
%& \qquad f_{2}(\cos\theta_{0}\cos\theta_{0}^{\prime}-\sin\theta_{0}\sin
%\theta_{0}^{\prime}\cos\theta_{1}^{\prime})\sin^{2d-2}\theta_{0}\sin
%^{2d-2}\theta_{0}^{\prime}\sin^{2d-3}\theta_{1}^{\prime}d\theta_{1}^{\prime
%}d\theta_{0}^{\prime}d\theta_{0}%
%\end{align*}

We shall work with the family of functions on $S^d$ ($d \geq 2$)
\begin{equation}\label{Base}
g_{k}(\theta_0)=\frac{1+\cos^{k}\theta_0}{V_{k}}, \theta_0 \in [0,\pi], k\in \N
\end{equation}
where
\[
V_k=\int_{S^{d}} g_k  dS{^d}
\]
The functions $g_k$ are a family of isotropic density functions on $S^d$ (as they are bounded and positive, and their integral is $1$). Furthermore, they are a base of the space of isotropic density functions. This can be proved using the Stone-Weierstrass Theorem ($\{1,x,x^2,\dotsc\}$ is a complete base on $L^2[-1,1]$), and deducing from it that $\{\cos^k(\theta)\,|\,k\geq 0 \}$ is a complete base on $L^2[0,\pi]$.

\begin{lem}\label{main}
For $d\geq2$, the following equality holds:%
\[
V(g_k+g_l)=\frac{2(V(g_k)+V(g_l))-V(g_k)V(g_l)}{2}
\]

\end{lem}
\begin{proof}
{Let us first compute the volume integrals%
\begin{align*}
V_{k}  & =\int_{S^d}\left(  1+\cos^{k}\theta_{0} \right)  dS^d%
=\left\vert S^{d-1}\right\vert \int_{0}^{\pi}\left(  1+\cos^{k}\theta
_{0}\right)  \sin^{d-1}\theta_{0}d\theta_{0}\\
& =\left\vert S^{d-1}\right\vert \left(  \int_{0}^{\pi}\sin^{d-1}\theta
_{0}d\theta_{0}+\int_{0}^{\pi}\cos^{k}\theta_{0}\sin^{d-1}\theta_{0}%
d\theta_{0}\right)
\end{align*}
that is,% HASTA AQUI
\begin{equation} \label{F1}
V_{k}=\left\{
\begin{array}
[c]{l}%
\left\vert S^{d-1}\right\vert K_d \frac{(d-2)!!}{(d-1)!!}\quad\text{for $k$ odd,} \\
\left\vert S^{d-1}\right\vert K_d \frac{(d-2)!!}{(d-1)!!}+\left\vert
S^{d-1}\right\vert K_d \frac{(k-1)!!(d-2)!!}{(d+k-1)!!}\quad\text{for $k$ even,}\\
\end{array}
\right.
\end{equation}
where $K_d=\pi$ for odd $d$, and $K_d=2$ for even $d$.

The individual variances are%
\begin{align*}
V(g_k)
& =2-2\frac{\left\vert S^{d-1}\right\vert }{V_{k}}\int_{0}^{\pi}\cos
\theta_{0}(1+\cos^{k}\theta_{0})\sin^{d-1}\theta_{0}d\theta_{0}\\
& =2-2\frac{\left\vert S^{d-1}\right\vert }{V_{k}}\int_{0}^{\pi}%
\cos^{k+1}\theta_{0}\sin^{d-1}\theta_{0}d\theta_{0}
\end{align*}
so that (after substitution of $V_k$ and using lemma \ref{intcossin})
\begin{equation}\label{F2}
V(g_k)=\left\{ \begin{array}[c]{l}
 2-2 \frac{k!!\, (d-1)!!}{(k+d)!!} \text{ for $k$ odd} \\
 2 \text{ for $k$ even}
 \end{array} \right.
\end{equation}
Hence, if $k$ or $l$ are even (for instance, assume $k$ even),%
\[
\frac{2(V(g_k)+V(g_l))-V(g_k)V(g_l)}{2}=\frac{2(2+V(g_l))-2V(g_l)}{2}=2,
\]
while for odd $k$ and $l$
\[
2(V(g_k)+V(g_l))=2\left(  4-2(d-1)!!\left(  \frac{k!!}{(k+d)!!}%
+\frac{l!!}{(l+d)!!}\right)  \right)  ,
\]
and also%
\begin{align*}
V(g_k)V(g_l)  & =\left(  2-2\frac{k!!(d-1)!!}{(k+d)!!}\right)  \left(
2-2\frac{l!!(d-1)!!}{(l+d)!!}\right) \\
& =4+4\frac{k!!\,l!!\,\left((d-1)!!\right)^{2}}
{(k+d)!!(l+d)!!}-4(d-1)!!\left(  \frac{k!!}{(k+d)!!}+\frac{l!!}{(l+d)!!}%
\right)  ;
\end{align*}
so, finally%
\begin{equation}\label{left}
\frac {2(V(g_k)+V(g_l))-V(g_k)V(g_l)}{2}
=2-2\frac{k!!\,l!!\left( (d-1)!!\right)^2}{(k+d)!!(l+d)!!}
\end{equation}

On the other hand, by corollary \ref{sum},
\[
V(g_k+g_l)=2-2\frac{|S^{d-2}||S^{d-1}| I_{k,l}}{V_k V_l},
\]
for
\begin{align*}
I_{k,l} =  & \int_{0}^{\pi}\int_{0}^{\pi}\int_{0}^{\pi}\cos\theta_{0}f_k%
(\cos\theta_{0}^{\prime})f_l(\cos\theta_{0}\cos\theta_{0}^{\prime}%
+\sin\theta_{0}\sin\theta_{0}^{\prime}\cos\theta_{1}^{\prime})\nonumber\\
& \cdot\sin^{d-1}\theta_{0}\sin^{d-1}\theta_{0}^{\prime}\sin^{d-2}%
\theta_{1}^{\prime}d\theta_{1}^{\prime}d\theta_{0}^{\prime}d\theta_{0}
\end{align*}
(where $f_k=g_k \circ \arccos$, that is, $g_k(\theta)=f_k(\cos\theta)$).

Developing the integral expression,
\begin{align}
I_{k,l}  &  =\int_{0}^{\pi}\int_{0}^{\pi}\int_{0}^{\pi}\cos\theta_{0}(1+\cos^{k}%
\theta_{0}^{\prime})(1+(\cos\theta_{0}\cos\theta_{0}^{\prime}+\sin\theta
_{0}\sin\theta_{0}^{\prime}\cos\theta_{1}^{\prime})^{l})\nonumber\\
& \cdot\sin^{d-1}\theta_{0}\sin^{d-1}\theta_{0}^{\prime}\sin^{d-2}%
\theta_{1}^{\prime}d\theta_{1}^{\prime}d\theta_{0}^{\prime}d\theta
_{0}\nonumber\\
& =\int_{0}^{\pi}\int_{0}^{\pi}\int_{0}^{\pi}\cos\theta_{0}\sin^{d-1}%
\theta_{0}\sin^{d-1}\theta_{0}^{\prime}\sin^{d-2}\theta_{1}^{\prime}%
d\theta_{1}^{\prime}d\theta_{0}^{\prime}d\theta_{0}\label{A}\\
& +\int_{0}^{\pi}\int_{0}^{\pi}\int_{0}^{\pi}\cos\theta_{0}\cos^{k}\theta
_{0}^{\prime}\sin^{d-1}\theta_{0}\sin^{d-1}\theta_{0}^{\prime}\sin
^{d-2}\theta_{1}^{\prime}d\theta_{1}^{\prime}d\theta_{0}^{\prime}d\theta
_{0}\label{B}\\
& +\int_{0}^{\pi}\int_{0}^{\pi}\int_{0}^{\pi}\cos\theta_{0}(1+\cos^{k}%
\theta_{0}^{\prime})(\cos\theta_{0}\cos\theta_{0}^{\prime}+\sin\theta_{0}%
\sin\theta_{0}^{\prime}\cos\theta_{1}^{\prime})^{l}\nonumber\\
& \cdot\sin^{d-1}\theta_{0}\sin^{d-1}\theta_{0}^{\prime}\sin^{d-2}%
\theta_{1}^{\prime}d\theta_{1}^{\prime}d\theta_{0}^{\prime}d\theta
_{0}\label{C}%
\end{align}
Integrals (\ref{A},\ref{B}) vanish by lemma \ref{intcossin}.

By using Newton binomial Theorem, integral (\ref{C}) becomes%
\begin{align}
& \!\! \int_{0}^{\pi}\!\!\! \int_{0}^{\pi}\!\!\! \int_{0}^{\pi}\!\!\!\cos\theta_{0}\!
(1+\cos^{k}\theta_{0}^{\prime})\left(  \sum_{h=0}^{l}\tbinom{l}{h}\cos
^{l-h}\theta_{0}\cos^{l-h}\theta_{0}^{\prime}\sin^{h}\theta_{0}\sin^{h}%
\theta_{0}^{\prime}\cos^{h}\theta_{1}^{\prime}\right) \!\! \nonumber\\
& \cdot\sin^{d-1}\theta_{0}\sin^{d-1}\theta_{0}^{\prime}\sin^{d-2}%
\theta_{1}^{\prime}d\theta_{1}^{\prime}d\theta_{0}^{\prime}d\theta
_{0}\nonumber\\
& =\sum_{h=0}^{l}\tbinom{l}{h} \nonumber \\
& \cdot \!\! \int_{0}^{\pi}\!\! \int_{0}^{\pi}
 \!\! \int_{0}^{\pi}
\cos^{l-h+1}\theta_{0}\cos^{l-h}\theta_{0}^{\prime}\cos^{h}%
\theta_{1}^{\prime}\sin^{d+h-1}\theta_{0}\sin^{d+h-1}\theta_{0}^{\prime}%
\sin^{d-2}\theta_{1}^{\prime}d\theta_{1}^{\prime}d\theta_{0}^{\prime}%
d\theta_{0}\label{E}\\
& +\sum_{h=0}^{l}\tbinom{l}{h} \nonumber \\
& \cdot \!\!\! \int_{0}^{\pi}\!\!\! \int_{0}^{\pi}\!\!\! \int
_{0}^{\pi}\!\!\! \cos^{l-h+1}\theta_{0}\cos^{l-h+k}\theta_{0}^{\prime}\cos^{h}%
\theta_{1}^{\prime}\sin^{d+h-1}\theta_{0}\sin^{d+h-1}\theta_{0}^{\prime}%
\sin^{d-2}\theta_{1}^{\prime}d\theta_{1}^{\prime}d\theta_{0}^{\prime}%
d\theta_{0}\label{F}%
\end{align}

Both in (\ref{E}) and in (\ref{F}) the terms with odd $h$ vanish
(since they are odd on $\cos\theta_{1}^{\prime}$). Furthermore, in (\ref{E})
the remaining terms also vanish, since the exponents of
$\cos\theta_{0}$ and $\cos\theta_{0}^{\prime}$ have different parity.
On the other hand, in the sum (\ref{F}), only the terms with even $h$, even $l-h-1$
(that is, odd $l$) and even $l-h+k$ ($\Rightarrow$ odd $k$) survive. Hence,
writing $h=2c$, and $k=2a+1,l=2b+1$ when $k,l$ are both odd:
\begin{align*}
I_{k,l}  & =0,\text{ for even }k\text{ or }l,\\
I_{2a+1,2b+1}  & =\sum_{c=0}^{b}\tbinom{2b+1}{2c}\int_{0}^{\pi}\int_{0}^{\pi
}\int_{0}^{\pi}\cos^{2(b-c+1)}\theta_{0}\cos^{2(a+b-c+1)}\theta_{0}^{\prime
}\cos^{2c}\theta_{1}^{\prime}\\
& \cdot\sin^{d+2c-1}\theta_{0}\sin^{d+2c-1}\theta_{0}^{\prime}\sin
^{d-2}\theta_{1}^{\prime}d\theta_{1}^{\prime}d\theta_{0}^{\prime}d\theta
_{0}\\
& =\sum_{c=0}^{b}\tbinom{2b+1}{2c}\int_{0}^{\pi}\cos^{2(b-c+1)}\theta_{0}%
\sin^{d+2c-1}\theta_{0}d\theta_{0}\\
& \cdot\int_{0}^{\pi}\cos^{2(a+b-c+1)}\theta_{0}^{\prime}\sin^{d+2c-1}%
\theta_{0}^{\prime}d\theta_{0}^{\prime}\int_{0}^{\pi}\cos^{2c}\theta
_{1}^{\prime}\sin^{d-2}\theta_{1}^{\prime}d\theta_{1}^{\prime},
\end{align*}
So, for even $k$, $l$, or both,
\[
V(g_k+g_l)=2=\frac{2(V(g_k)+V(g_l))-V(g_k)V(g_l)}{2},
\]
and only the case $k=2a+1$, $l=2b+1$ remains to be proved.

In this case,
\begin{align*}
I_{2a+1,2b+1}  & =\sum_{c=0}^{b}\left(  \frac{(2b+1)!}{(2c)!(2(b-c)+1)!}%
K_d\frac{(2(b-c)+1)!!(2c+d-2)!!}{(2b+d+1))!!}\right. \\
& \left.  \cdot K_d\frac{(2(a+b-c)+1)!!(2c+d-2)!!}{(2(a+b)+d+1)!!}%
K_{d-1} \frac{(2c-1)!!(d-3)!!}{(2c+d-2)!!}\right) \\
& =K_d^{2}K_{d-1}\frac{(2b+1)!(d-3)!!}{(2b+d+1)!!(2(a+b)+d+1)!!} \\
& \cdot\sum_{c=0}^{b}%
\frac{(2(b-c)+1)!!(2c+d-2)!!(2(a+b-c)+1)!!(2c-1)!!}{(2c)!(2(b-c)+1)!}\\
& =K_d^{2}K_{d-1}\frac{(2b+1)!(d-3)!!}{(2b+d+1)!!(2(a+b)+d+1)!!} \\
& \quad \cdot \sum_{c=0}^{b}%
\frac{(2c+d-2)!!(2(a+b-c)+1)!!}{(2c)!!(2(b-c))!!}.%
\end{align*}

Now
\[
V(g_k+g_l)=2-2\frac{\left\vert S^{d-1}\right\vert \left\vert S^{d-2}%
\right\vert }{V_{2a+1}V_{2b+1}}I_{2a+1,2b+1},
\]
and
\begin{align*}
\frac{\left\vert S^{d-1}\right\vert \left\vert S^{d-2}\right\vert
}{V_{2a+1}V_{2b+1}}
&=\frac{\left\vert S^{d-1}\right\vert \left\vert S^{d-2}\right\vert
}{K_d^2 \left\vert S^{d-1}\right\vert ^2 \frac{((d-2)!!)^2}{((d-1)!!)^2}}\\
&=\frac{\left\vert S^{d-2}\right\vert ((d-1)!!)^2}{K_d^2 \left\vert S^{d-1}\right\vert ((d-2)!!)^2}
\end{align*}

Notice that for even $d=2p$
\[
\frac{\left\vert S^{d-2}\right\vert}{\left\vert S^{d-1}\right\vert }
=\frac{\left\vert S^{2p-2}\right\vert}{\left\vert S^{2p-1}\right\vert}
= \frac{(2p-2)!!}{\left(  2\pi\right)  ^{p}}\frac{2\left(  2\pi\right)  ^{p-1}}{(2p-3)!!}
= \frac{1}{\pi}\frac{(d-2)!!}{(d-3)!!}
= \frac{1}{K_{d-1}}\frac{(d-2)!!}{(d-3)!!}
\]
while for odd $d=2p+1$
\[
\frac{\left\vert S^{d-2}\right\vert}{\left\vert S^{d-1}\right\vert }
=\frac{\left\vert S^{2p-1}\right\vert}{\left\vert S^{2p}\right\vert}
= \frac{(2p-1)!!}{2\left(  2\pi\right)  ^{p}}\frac{\left(  2\pi\right)  ^{p}}{(2p-2)!!}
= \frac{1}{2}\frac{(d-2)!!}{(d-3)!!}
= \frac{1}{K_{d-1}} \frac{(d-2)!!}{(d-3)!!}
\]
Substitution yields
\[
V(g_k+g_l)
=2-2\frac{ ((d-1)!!)^2}{K_d^2 K_{d-1} (d-2)!!\,(d-3)!!}I_{2a+1,2b+1},
\]
that is,
\begin{align}
V(g_k+g_l)&=2-2\frac{((d-1)!!)^2(2b+1)!}{(d-2)!!(2b+d+1)!!(2(a+b)+d+1)!!}  \nonumber \\
&\qquad \cdot \sum_{c=0}^{b}%
\frac{(2c+d-2)!!(2(a+b-c)+1)!!}{(2c)!!(2(b-c))!!}. \label{F4}
\end{align}
And we have to check that the result above is equal to
\begin{align*}
\frac {2(V(g_{2a+1})+V(g_{2b+1}))-V(g_{2a+1})V(g_{2b+1})}{2}
\\
=2-2\frac{(2a+1)!!\,(2b+1)!!\left( (d-1)!!\right)^2}{(2a+d+1)!!(2b+d+1)!!}
\end{align*}
So, it suffices to check that
\begin{multline*}
\frac{((d-1)!!)^2(2b+1)!}{(d-2)!!(2b+d+1)!!(2(a+b)+d+1)!!}\sum_{c=0}^{b}%
\frac{(2c+d-\! 2\! )!!(2(a+b-c)+\! 1\! )!!}{(2c)!!(2(b-c))!!}
\\ =\frac{(2a+1)!!\,(2b+1)!!\left( (d-1)!!\right)^2}{(2a+d+1)!!(2b+d+1)!!},
\end{multline*}
or, after simplification,
\[
\frac{(2b)!!}{(d-2)!!(2(a+b)+d+1))!!}\sum_{c=0}^{b}\frac
{(2c+d-2)!!(2(a+b-c)+1)!!}{(2c)!!(2(b-c))!!}
=\frac{(2a+1)!!}{(2a+d+1)!!},
\]
that is,
\[
\sum_{c=0}^{b}\frac{(2c+d-2)!!(2(a+b-c)+1)!!}{(2c)!!(2(b-c))!!}%
=\frac{(2a+1)!!(d-2)!!(2(a+b)+d+1)!!}{(2b)!!(2a+d+1)!!}
\]

To find out the explicit expression of the sum, we can use%
\[
(2c)!!(2(b-c))!!=2^{b}c!(b-c)!=2^{b}b!\frac{c!(b-c)!}{b!}=\frac{(2b)!!}%
{\tbinom{b}{c}}
\]
to obtain%
\begin{align*}
& \sum_{c=0}^{b}\frac{(2c+d-2)!!(2(a+b-c)+1)!!}{(2c)!!(2(b-c))!!}\\
& =\frac{1}{(2b)!!}\sum_{c=0}^{b}\tbinom{b}{c}(2c+d-2)!!(2(a+b-c)+1)!!
\end{align*}
Therefore, to prove the result it suffices to check that%
\[
\sum_{c=0}^{b}\tbinom{b}{c}(2c+d-2)!!(2(a+b-c)+1)!!=\frac
{(2a+1)!!(d-2)!!(2(a+b)+d+1)!!}{(2a+d+1)!!}.
\]
We prove this fact in the following lemma, thus finishing the proof}% \hfill$\Box$
\end{proof}

\begin{remark} \label{rmk34}
The proof for $d=1$ is analogous, using remark~\ref{rmk25} instead of
corollary~\ref{sum}. Taking into account the identities $0!!=(-1)!!=1$ and
$|S^0|=2$, the formulas~(\ref{F1}), (\ref{F2}) and (\ref{left}) are readily
checked to hold. Also, the integral $I_{k,l}$ vanishes for odd $k$ or $l$ and, finally, the formula~(\ref{F4}) holds. From this point, the proof for $d=1$ is identical to the general case.
\end{remark}

\begin{lem}
The following equality holds:
\[
\sum_{c=0}^{b}\tbinom{b}{c}(2c+d-2)!!(2(a+b-c)+1)!!=\frac
{(2a+1)!!(d-2)!!(2(a+b)+d+1)!!}{(2a+d+1)!!}.
\]
\end{lem}
\begin{proof}
{For odd $d=2p-1$, the proof is based on counting increasing ordered rooted trees. Callan \cite{Cal} shows that there are $(2n-1)!!$ increasing ordered trees with $n+1$ vertices. Let us now consider the set of increasing ordered rooted trees with basic subtrees of size $p$ and $a+2$ (including the root).

\begin{figure}[htb]
\begin{center}
\includegraphics[width=0.3\textwidth]{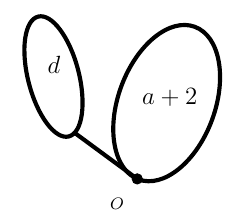}
\end{center}
\fcaption{An ordered tree with basic subtrees of size $p$ and $a+2$}
\end{figure}

Now, let us add $b$ vertices (with greater labels) to the basic subtrees. In general, to add a greater label to a fixed ordered tree of size $n$, there are $2n-1$ choices.
Hence, for the first additional vertex, we have $(2p-1)+(2(a+2)-1)=2(p+a)+2$ possible choices. In a similar way, there are $2(p+a)+4$ choices for the second vertex,
$2(p+a)+6$ for the third, and $2(p+a)+2b$ for the last one. So, the total number of ordered trees
constructed in this fashion is
\[
(2p-3)!!(2a+1)!!\frac{(2(p+a+b))!!}{(2(p+a))!!}
\]
That is,
\[
(d-2)!!(2a+1)!!\frac{(2(a+b)+d+1)!!}{(2a+d+1))!!}
\]

We can also count the number of elements of this family of ordered trees in a different way.
If we add exactly $c$ vertices to the left basic subtree (there are $\binom{b}{c}$ choices of vertices),
 and the remaining $b-c$ vertices to the right subtree, there are $(2(p+c)-3)!!$ choices for the left subtree
 and $(2(a+b-c)+1)!!$ for the right subtree. So, the number of trees is
\[
\sum_{c=0}^b \binom{b}{c}(2(p+c)-3)!!(2(a+b-c)+1)!!=\sum_{c=0}^b \binom{b}{c}(2c+d-2)!!(2(a+b-c)+1)!!
\]
and thus the proof is complete for odd values of $d$.

For $d=2p$, we have to prove that
\[
\sum_{c=0}^{b}\dbinom{b}{c}\frac{(2(c+p-1))!!}{(2(p-1))!!}(2(a+b-c)+1)!!
=\frac {(2a+1)!!(2(a+b+p)+1)!!}{(2(a+p)+1)!!}
\]
In a similar way to the odd case, the right hand side counts the number of configurations for an increasing ordered rooted tree built from a fixed left subtree $T$ of size $p$ and a basic right subtree of size $a+2$, after adding $b$ vertices with greater labels than those on the basic subtrees. In this way, there are $2(p+a+2)-1=2(a+p)+3$ choices to place the first vertex, $2(p+a+3)-1=2(a+p)+5$ for the second one, and $2(p+a+2+b-1)-1=2(a+p+b)+1$ choices for the last vertex.

The left hand side can be seen as a different way of counting the same family. We choose $c$ from the $b$ vertices (can be done in $b$ over $c$ ways); add the remaining $b-c$ vertices to the right subtree (this gives $(2(a+b-c)+1)!!$ configurations); then we have to add the $c$ chosen vertices to left subtree of size $p+1$ (this time counting also the root of the global tree), but taking into account that the rightmost son of the root must be one of the $p$ original vertices of $T$ (to avoid counting some configurations twice). The first vertex can thus be added in $2(p+1)-1-1=2p$ ways; in a similar way, there are $2p+2$ choices for the second vertex, and $2(p+c-1)$ choices for the last one. Summing over all possible values of $c$, from $0$ to $b$, completes the proof}%. \hfill$\Box$
\end{proof}

\begin{thm}\label{mainthm}
Given two independent isotropic r.vs. in $S^d$, $X_1$ and $X_2$, then it holds that
$$
V(X_1+X_2)=V(X_1)+V(X_2)-\frac{V(X_1)V(X_2)}{2}
$$
\end{thm}
\begin{proof}
{Let us decompose the density functions of $X_1$ and $X_2$, $f_1$ and $f_2$ respectively, into infinite linear combinations of density functions of the base $B=\{\,g_k\ |\ k\in\N^+\}$ of $L^2[0,\pi]$:
$$
f_1=\sum_{j=0}^\infty\alpha_j\,g_j
\quad\text{and}\quad
f_2=\sum_{k=0}^\infty\beta_k\,g_k
\quad\text{such that}\quad
\sum_{j=0}^\infty\alpha_j = \sum_{k=0}^\infty\beta_k=1
$$
Then, applying Lemma~\ref{main} and the previous equalities we get:
$$
\begin{array}{lcl}
V(X_1+X_2) & = & \displaystyle\sum_{j,k}\alpha_j\,\beta_k\,V(X_{g_j}+X_{g_k}) \\
& = & \displaystyle\sum_{j,k}\alpha_j\,\beta_k\left(V(X_{g_j})+V(X_{g_k})-\frac{V(X_{g_j})\,V(X_{g_k})}{2}\right) \\
& = & \displaystyle\sum_{k}\beta_k\,V(X_1)+\sum_{j}\alpha_j\,V(X_2)-\frac{1}{2}\,\sum_{j,k}\alpha_j\,\beta_k\,V(X_{g_j})\,V(X_{g_k}) \\
& = & \displaystyle V(X_1)+V(X_2)-\frac{1}{2}\,\left(\sum_j\alpha_j\,V(X_{g_j})\right)\,\left(\sum_k\beta_k\,V(X_{g_k})\right) \\
& = & \displaystyle V(X_1)+V(X_2)-\left.\frac{\displaystyle V(X_1)\,V(X_2)}{\displaystyle 2}\right. \\
\end{array}
$$
}\end{proof}

\vspace*{1pt}\textlineskip

\subsection{The general case}

We first prove that theorem \ref{mainthm} also holds for general r.vs. on $S^1$.
\begin{thm}
Given two independent r.vs. in $S^1$, $X_1$ and $X_2$, then it holds that
$$
V(X_1+X_2)=V(X_1)+V(X_2)-\frac{V(X_1)V(X_2)}{2}
$$
\end{thm}
\begin{proof}
{It suffices to prove lemma~\ref{main} for general r.vs. on $S^1$. From there, the proof would be completely similar to that of theorem \ref{mainthm}}%. \hfill$\Box$
\end{proof}

In order to extend lemma~\ref{main} to $S^1$, we need to extend the base given by functions (\ref{Base}) on $ [0,\pi ]$ to a base on $[ -\pi,\pi ]$ by adding
\[
h_k(\theta_0)=\frac{1+\sin^k(\theta_0)}{V_k}, \qquad \theta_0\in [-\pi,\pi ],
    \,k\in\mathbb{N}^*,
\]
where $V_K$ is the same normalization constant of the functions $g_k$.

\begin{lem} For $d=1$ the following equality holds:
\[
V(f_k+f_l)=\frac{2(V(f_k)+V(f_l))-V(f_k)V(f_l)}{2},
\]
where $f_k,f_l\in \{g_{k'}\,|\,k'\in\mathbb{N}\}\cup \{h_{k'}\,|\,k'\in\mathbb{N}^*\}$.
\end{lem}
\begin{proof}
{Note that functions $g_k(\theta_0),\,k\in\mathbb{N}$ are isotropic (\emph{i.e.} symmetric) while functions $h_k(\theta_0),\,k\in\mathbb{N}^*$ are not.
We distinguish three cases:
\begin{description}
\item[i)] $f_k=g_k$ and $f_l=g_l$. This case is already proved in remark~\ref{rmk34}.
\item[ii)] $f_k=g_k$ and $f_l=h_l$.
\item[iii)]$f_k=h_k$ and $f_l=g_l$.
\end{description}
These last two cases are straightforward, taking into account that $V(h_k)=2$ for all $k\in\mathbb{N}^*$ and hence
\[
    \frac{2(V(f_k)+V(f_l))-V(f_k)V(f_l)}{2}=2.
\]
On the other hand, the density function of the sum is
\[
f(\theta_0)=\int_{-\pi}^\pi f_k(\alpha)h_l(\beta') d\theta'_0,
\]
where $\alpha=\cos(\theta'_0)$ and $\beta'=\cos(\theta_0)\sin(\theta'_0)+ \sin(\theta_0)\cos(\theta'_0)$, since $h_l$ depends on the sine function and not on the cosine. When computing the variance of the sum, all the integrals of trigonometric functions vanish identically, so that only $E(2-2\cos(\theta))=E[2]=2$ remains, and the proof is complete}
%\hfill$\Box$
\end{proof}

In the general case $d\geq 2$, we conjecture that Theorem \ref{mainthm} holds for general r.vs. on $S^d$. We illustrate this fact with an example.

\begin{example}
Let us consider the density functions on $S^7$:
\begin{align*}
g_1&=\frac{2+\cos(\theta_0)\sin^2(\theta_1)+\cos(\theta_1)\sin(\theta_2)}{V_1},\\
g_2&=\frac{1+\cos(\theta_0)}{V_2}.
\end{align*}
For these functions, the volumes are
\begin{align*}
V_1= \frac{2}{3}\pi^4,\qquad V_2=\frac{1}{3}\pi^4.
\end{align*}
Computation of individual variances yields
\begin{align*}
V(g_2)&=7/4, \\
V(g_1)&=2-2\|S^4\|\int_0^\pi\int_0^\pi\int_0^\pi\cos(\theta_0)g_1(\theta_0,\theta_1,\theta_2)
\\
&\qquad \qquad\ \cdot sin^6(\theta_0)\sin^5(\theta_1)\sin^4(\theta2)d\theta_0d\theta_1d\theta_2
=\frac{53}{28}
\end{align*}
Hence,
\[
\frac{2(g_1)+V(g_2))-V(g_1)V(g_2)}{2}=\frac{445}{224}
\]
On the other hand, a reasoning similar to the proof of corollary \ref{sum} allows to compute the variance of the sum as
\begin{align*}
V(g_1+g_2)=&2-2\|S^4\|^2\int_0^\pi\int_0^\pi\int_0^\pi\int_0^\pi\int_0^\pi\int_0^\pi
\cos(\theta_0) g_1(\theta_0 ^\prime,\theta_1 ^\prime,\theta_2 ^\prime) g_2(\beta)  \\
&\qquad \cdot
\sin^6(\theta_0)\sin^5(\theta_1)\sin^4(\theta_2)
\sin^6(\theta_0^\prime)\sin^5(\theta_1^\prime)\sin^4(\theta_2^\prime) \\
&\qquad \qquad d\theta_0 d\theta_1 d\theta_2
d\theta_0^\prime d\theta_1^\prime d\theta_2^\prime,
\end{align*}
where
\begin{multline*}
\beta=\cos(\theta_0)\cos(\theta_0 ^\prime)+\sin(\theta_0)\sin(\theta_0 ^\prime)\cos(\theta_1)\cos(\theta_1 ^\prime) \\
+\sin(\theta_0)\sin(\theta_0 ^\prime)\sin(\theta_1)\sin(\theta_1 ^\prime)\cos(\theta_2)\cos(\theta_2 ^\prime).
\end{multline*}
Computation of the integral yields $445/224$, so that the formula holds in this case.
\end{example}

\vspace*{1pt}\textlineskip

\section{Conclusions}

\vspace*{1pt}\textlineskip

\subsection{Analysis of the accumulated error distribution}

First we generalize formula (\ref{VarFormula}) for an arbitrary number $k$ of independent quantum computing errors. These errors can be generated during the execution of a quantum computation, one in each time interval corresponding to a logical block into which the quantum algorithm is divided. Since errors correspond to different time intervals, there is no reason not to assume that they are independent because, as explained in the following subsection, they are generated by errors in the application of quantum gates and by the random interaction with the environment. The obtained formula can be applied for isotropic errors and, for general errors, if we assume that conjecture \ref{Conjetura} is true. If all errors have the same variance $\sigma$, the formula can easily be demonstrated by induction from theorem \ref{mainthm}.
$$
V(E_1+\cdots+E_k)=2-2\left(1-\frac{\sigma}{2}\right)^k
$$
The proof can be easily generalized for different variances $\sigma_1,\dots,\sigma_k$.
$$
V(E_1+\cdots+E_k)=\sum_{j=1}^{k} (-1)^{j+1}\frac{s_j}{2^{j-1}}
\quad\text{where}\quad
\begin{array}{l}
s_1=\sigma_1+\sigma_2+\cdots+\sigma_k \\
s_2=\sigma_1\sigma_2+\sigma_1\sigma_3+\cdots+\sigma_{k-1}\sigma_k \\
\cdots \\
s_k=\sigma_1\sigma_2\cdots\sigma_k
\end{array}
$$

\begin{figure}[htb]
\begin{center}
\includegraphics[width=\textwidth]{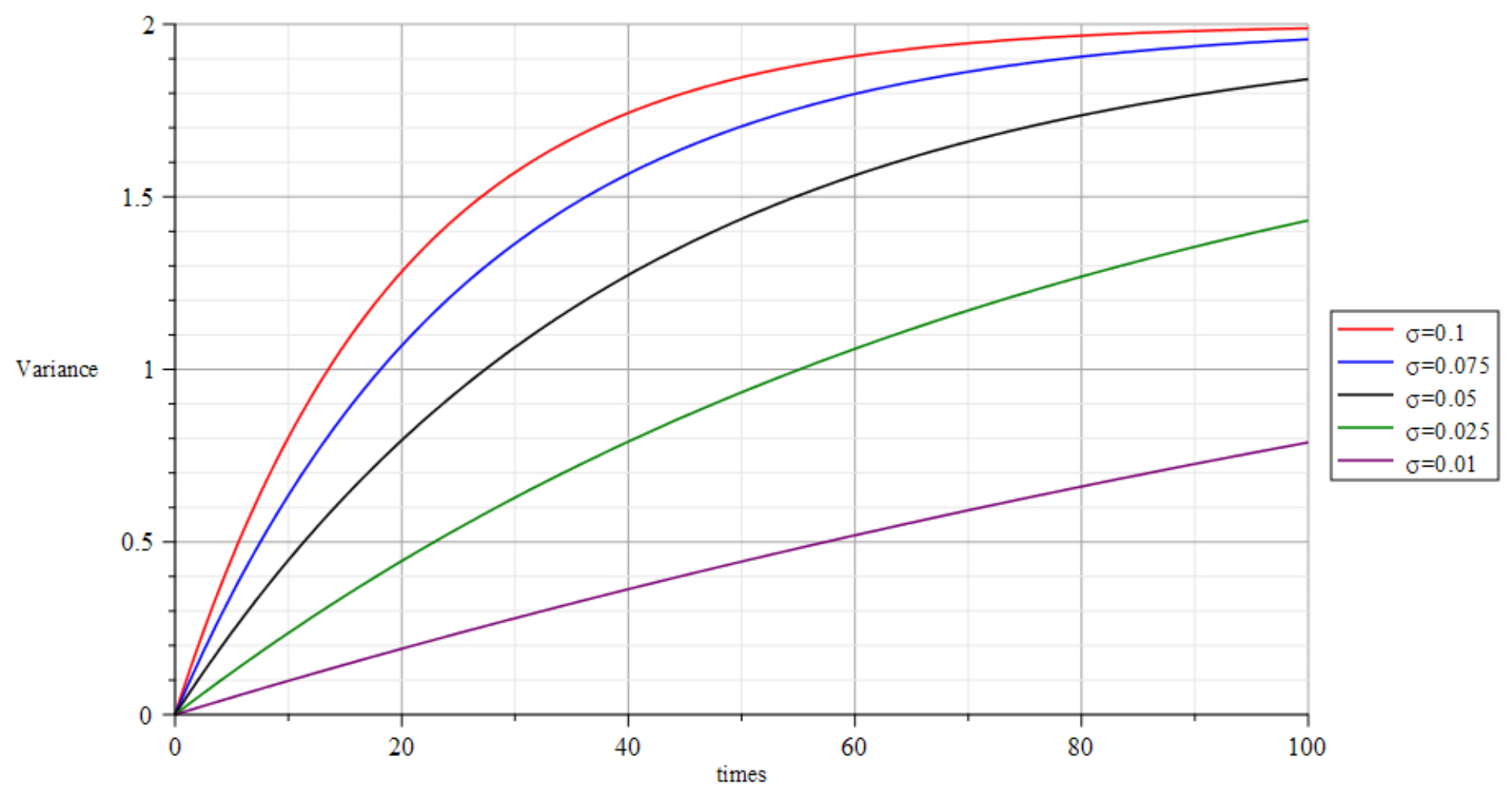}
\end{center}
\fcaption{Error behavior for $k=100$ and $\sigma=0.1,\ 0.075,\ 0.05,\ 0.025$ and $0.01$}
\label{Errores}
\end{figure}

For all $0<\sigma<2$ it is verified that the variance of the accumulated quantum computing error tends to $2$ when the number of errors $k$ tends to infinity (see figure~\ref{Errores}).

In addition, if we want, for example, that the variance of the accumulated error was less than or equal to a constant $\sigma_{max}$, the variance of each error must verify
\begin{eqnarray}\label{VarInicial}
\sigma<\dfrac{C(\sigma_{max})}{k}
\end{eqnarray}
where $C(\sigma_{max})$ is a constant that depends on $\sigma_{max}$, for example $C(1)\approx 4/3$. Therefore, to achieve this goal it is necessary that the variance of quantum computing errors is smaller than $C(\sigma_{max})/k$, i.e. smaller than $C(\sigma_{max})$ times the inverse of the algorithm execution time.

Formula~(\ref{VarInicial}) establishes a threshold to the variance of quantum computing errors, that depends inversely on computation time, and connects the results of this article with the threshold theorem~\cite{AGP,SDT,WFSH,CT,WFHH}, key in the study of the feasibility of quantum computing.

In addition to the variance, it is important to analyze the shape of the density function, particularly the dimension of its support (set of points in the domain where the density function is greater than zero). For example, the support of a density function that represents the error in a qubit has dimension $4$, far from the support dimension of an isotropic error that can be $2^{n-1}-1$. The dimension of the support of the sum of two independent errors is less than or equal to the sum of the dimensions of the respective supports. A well-known limitation on the growth of the accumulated error support dimension is obtained considering local errors. Thus, for example, the support of independent errors in each qubit will have a maximum dimension of $4n$.

Another important aspect to analyze in the accumulated error density function is the form of this function on its support. It is expected, in the same line as the central limit theorem, that the density function tends to be uniform over its support. This is the case, for example, if we accumulate independent isotropic errors equally distributed with density functions given by formula~(\ref{Base}).

\vspace*{1pt}\textlineskip

\subsection{Types of quantum computing errors}

Quantum computing must deal with two types of errors: quantum gates errors and decoherence. In the first one, we can also include measurement errors and those of preparation of initial qubits states. Despite being local errors that affect one or two qubits, their control is an important challenge for quantum computing. When we use quantum error correction codes to control these errors, there are always two issues that prevent a perfect control. First, we do not have codes that correct any error of one or two qubits and that subtly encode all the qubits involved in a quantum computation. Therefore, quantum gates that involve qubits whose simultaneous errors are not controlled by the code are not protected. Second, once the algorithm for the coded version has been expanded, it is necessary to apply the code correction circuit after each of the quantum gates of the algorithm. Therefore the quantum gates of these circuits (those corresponding to the last code, if we are using concatenated codes) remain unprotected. In addition, the number of gates of these correction circuits is much greater than the number of gates of the original algorithm. Errors not controlled by these circumstances accumulate during computation and we can apply the results introduced in this article for their analysis.

Quantum decoherence is a much less known source of quantum computing error. It affects all qubits simultaneously and for this reason it cannot be controlled by quantum error correction codes. It is caused by the interaction of the $n-$qubit with the environment. The simplest decoherence model assumes that the errors are local in each of the qubits. However, the loss of entanglement requires the decoherence to be global in the set of entangled qubits. This fact makes decoherence more difficult to control.

In both cases the analysis presented in this article is applicable. In the first, the error accumulated by decoherence will be global, although its density function will have a support with small dimension (a maximum of $4n$, as we have seen in the previous subsection). In the second, probably more realistic, the support of the density function of the accumulated error by decoherence can have maximum dimension, $2^{n-1}-1$, if the algorithm uses highly entangled $n-$qubits.

In summary, both quantum gate errors and decoherence errors cannot be completely controlled by quantum error correction codes. In particular, these are not able at all to control isotropic errors~\cite{LPF} and it is foreseeable that the same will happen to errors that affect all the qubits simultaneously. Consequently, the formula~(\ref{VarInicial}) will be important to analyze the viability of any quantum computation.


\begin{thebibliography}{99}
\bibitem{Ga} Gaitan, F., Quantum error correction and fault tolerant quantum computing, {\sl CRC Press}, 2008.
\bibitem{CS} Calderbank, A.R., Shor, P.W., Good quantum error-correcting codes exist, {\sl Phys. Rev. A} {\bf 54}, 1098--1105, (1996).
\bibitem{St1} Steane, A.M., Multiple particle inference and quantum error correction, {\sl Proc. Roy. Soc. A.} {\bf 452} 2551, (1996).
\bibitem{Go1} Gottesman, D., Stabilizer Codes and Quantum Error Correction. {\sl PhD thesis}, California Institute of Technology, 1997.
\bibitem{Sh} Shor, P.W., Fault-tolerant quantum computation. {\sl Symposium on the Foundations of Computer Science}, Los Alamitos, CA, (1996).
\bibitem{Pr1} Preskill, J., Reliable quantum computers, {\sl Proc. Roy. Soc. Lond. A} {\bf 454}, 385--410, (1998).
\bibitem{St2} Steane, A.M., Active stabilization, quantum computation and quantum state synthesis, {\sl Phys. Rev. Lett.} {\bf 78}, 2252 (1997).
\bibitem{Go2} Gottesman, D., Theory of fault-tolerant quantum computation, {\sl Phys. Rev. A} {\bf 57}, 127--137 (1998).
\bibitem{NC} Nielsen, M.A.;  Chuang, I.L., Quantum Computation and Quantum Information, {\sl Cambridge University Press}, 2010.
\bibitem{LPF} Lacalle, J., Pozo Coronado, L.M., Fonseca de Oliveira, A.L., Isotropic errors and quantum correction codes, {\sl preprint} (2019).
\bibitem{Bie} Bienaym\'e, I.-J., Consid\'erations \`a l'appui de la d\'ecouverte de Laplace sur la loi de probabilit\'e dans la m\'ethode des moindres carr\'es, {\sl Comptes rendus de l'Acad\'emie des sciences Paris} {\bf 37},~309--317~(1853).
\bibitem{Pr2} Preskill, J., Sufficient condition on noise correlations for scalable quantum computing, {\sl Quantum Information \& Computation} {\bf 13}(3-4), 181--194 (2013).
\bibitem{Go3} Gottesman, D., Fault-tolerant quantum computation with constant overhead, {\sl Quantum Information \& Computation} {\bf 14}(15), 1338--1372 (2014).
\bibitem{CDT} Cross, A.W., Divincenzo, D.P., Terhal, B.M., A comparative code study for quantum fault tolerance, {\sl Quantum Information \& Computation} {\bf 9}(7), 541--572 (2009).
\bibitem{PR} Paetznick, A., Reichardt, B.W., Fault-tolerant ancilla preparation and noise threshold lower bounds for the 23-qubit Golay code, {\sl Quantum Information \& Computation} {\bf 12}(11), 1034--1080 (2012).
\bibitem{HFWH} Hill, C.D., Fowler, A.G., Wang, D.S., Hollenberg, L.C.L., Fault-tolerant quantum error correction code conversion, {\sl Quantum Information \& Computation} {\bf 13}(5), 439--451 (2013).
\bibitem{AZ} Ahsan, M., Zilqurnain Naqvi, S.A., Performance of topological quantum error correction in the presence of correlated noise, {\sl Quantum Information \& Computation} {\bf 18}(9), 743--778 (2018).
\bibitem{DP} Duclos-Cianci, G., Poulin, D., Fault-tolerant renormalization group decoder for abelian topological codes, {\sl Quantum Information \& Computation} {\bf 14}(9), 721--740 (2014).
\bibitem{Fo} Fowler, A.G., Minimum weight perfect matching of fault-tolerant topological quantum error correction in average O(1) parallel time, {\sl Quantum Information \& Computation} {\bf 15}(1), 145--158 (2015).
\bibitem{PFW} Paler,A., Fowler, A.G., Wille, R., Online scheduled execution of quantum circuits protected by surface codes, {\sl Quantum Information \& Computation} {\bf 17}(15), 1335--1348 (2017).
\bibitem{AGP} Aliferis, P., Gottesman, D., Preskill, J., Quantum accuracy threshold for concatenated distance-3 codes, {\sl Quantum Information \& Computation} {\bf 6}, 97--165 (2006).
\bibitem{SDT} Svore, K.M., DiVincenzo, D.P., Terhal, B., Noise threshold for a fault-tolerant two-dimensional lattice architecture, {\sl Quantum Information \& Computation} {\bf 7}(4), 297--318 (2007).
\bibitem{WFSH} Wang, D.S., Fowler, A.G., Stephens, A.M., Hollenberg, and L.C.L., Threshold error rates for the toric and planar codes, {\sl Quantum Information \& Computation} {\bf 10}(5), 456--469 (2010).
\bibitem{CT} Criger, B., Terhal, B., Noise thresholds for the [4,2,2]-concatenated toric code, {\sl Quantum Information \& Computation} {\bf 16}(15) 1261--1281 (2016).
\bibitem{WFHH} Wang, D.S., Fowler, A.G., Hill, C.D., Hollenberg, L.C.L., Graphical algorithms and threshold error rates for the 2d color code, {\sl Quantum Information \& Computation} {\bf 10}(9), 780--802 (2010).
\bibitem{OG} Ozen, M., Guzeltepe, M., Quantum codes from codes over Gaussian integers with respect to the Mannheim metric, {\sl Quantum Information \& Computation} {\bf 12}(9), 813--819 (2012).
\bibitem{LZLX} Li, R., Zou, F., Liu, Y., Xu, Z., Hermitian dual containing BCH codes and Construction of new quantum codes, {\sl Quantum Information \& Computation} {\bf 13}(1), 21--35 (2013).
\bibitem{SS} Sari, M., Siap, I., Quantum codes over a class of finite chain ring, {\sl Quantum Information \& Computation} {\bf 16}(1), 39--49 (2016).
\bibitem{Ha1} Hastings, M.B., Weight reduction for quantum codes, {\sl Quantum Information \& Computation} {\bf 17}(15), 1307--1334 (2017).
\bibitem{Fu} Fujita, H., Some classes of concatenated quantum codes: constructions and lower bounds, {\sl Quantum Information \& Computation} {\bf 15}(5), 385--405 (2015).
\bibitem{DPB} Dias de Albuquerque, C., Palazzo Jr., R., Brandani da Sil, E., Families of codes of topological quantum codes from tessellations tessellations \{4i+2,2i+1\}, \{4i,4i\}, \{8i-4,4\} and \{12i-6,3\}, {\sl Quantum Information \& Computation} {\bf 14}(15), 1424--1440 (2014).
\bibitem{GM} Greenbaum, D., Dutton, Z., Modeling coherent errors in quantum error correction, {\sl Quantum Science and Technology} {\bf 3}(1), pp. 015007 (2018).
\bibitem{BEKP} Bravyi, S., Englbrecht, M., Konig, R., Peard N., Correcting coherent errors with surface codes, {\sl npj Quantum Information} {\bf 4}, art. 55 (2018).
\bibitem{PSVW} Piltz, C., Sriarunothai, T., Var\'on, A.F., Wunderlich, C., A trapped-ion-based quantum byte with $10^{-5}$ next-neighbour cross-talk, {\sl Nature Communications} {\bf 5}, art. 4679 (2014).
\bibitem{BTS} Buterakos, D., Throckmorton, R.E., Das Sarma, S., Crosstalk error correction through dynamical decoupling of single-qubit gates in capacitively coupled singlet-triplet semiconductor spin qubits, {\sl Phys. Rev. B} {\bf 97}(4), pp. 045431 (2018).
\bibitem{Zu} Zurek, W.H., Decoherence and the Transition from Quantum to Classical - Revisited. In: Duplantier B., Raimond JM., Rivasseau V. (eds) Quantum Decoherence. Progress in Mathematical Physics, vol 48. Birkh\"auser Basel (2006).
\bibitem{Cal} Callan, D., A Combinatorial Survey of Identities for the Double Factorial, {\sl arXiv:0906.1317} (2009).
\end{thebibliography}
\end{document}